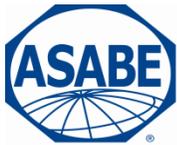

2950 Niles Road, St. Joseph, MI 49085-9659, USA
269.429.0300  fax 269.429.3852  hq@asabe.org  www.asabe.org

*An ASABE Meeting Presentation*
*DOI: https://doi.org/10.13031/aim.20 2301158*
*Paper Number: 2301158*# A Web-Based Application Leveraging Geospatial Information to Automate On-Farm Trial Design

Sneha Jha, Yaguang Zhang, James V. Krogmeier and Dennis R. Buckmaster**Written for presentation at the**
**2023 ASABE Annual International Meeting**
**Sponsored by ASABE**
**Omaha, Nebraska**
**July 9-12, 2023**

**ABSTRACT.** *On-farm sensor data have allowed farmers to implement field management techniques and intensively track the corresponding responses. These data combined with historical records open the door for real-time field management improvements with the help of current advancements in computing power. However, despite these advances, the statistical design of experiments is rarely used to evaluate the performance of field management techniques accurately. Traditionally, randomized block design is prevalent in statistical designs of field trials, but in practice it is limited in dealing with large variations in soil classes, management practices, and crop varieties. More specifically, although this experimental design is suited for most trial types, it is not the optimal choice when multiple factors are tested over multifarious natural variations in farms, due to the economic constraints caused by the sheer number of variables involved. Experimental refinement is required to better estimate the effects of the primary factor in the presence of auxiliary factors. In this way, farmers can better understand the characteristics and limitations of the primary factor. This work presents a framework for automating the analysis of local field variations by fusing soil classification data and lidar topography data with historical yield. This framework will be leveraged to automate designing of field experiments based on multiple topographic features*

**Keywords.** *[Click here to enter keywords and key phrases, separated by commas, with a period at the end]*The authors are solely responsible for the content of this meeting presentation. The presentation does not necessarily reflect the official position of the American Society of Agricultural and Biological Engineers (ASABE), and its printing and distribution does not constitute an endorsement of views which may be expressed. Meeting presentations are not subject to the formal peer review process by ASABE editorial committees; therefore, they are not to be presented as refereed publications. Publish your paper in our journal after successfully completing the peer review process. See www.asabe.org/JournalSubmission for details. Citation of this work should state that it is from an ASABE meeting paper. EXAMPLE: Author's Last Name, Initials. 2023. Title of presentation. ASABE Paper No. ---. St. Joseph, MI.: ASABE. For information about securing permission to reprint or reproduce a meeting presentation, please contact ASABE at www.asabe.org/copyright (2950 Niles Road, St. Joseph, MI 49085-9659 USA).## 1   Introduction

Agricultural research aims to use statistical designs of experiments to analyze the significance of treatment effects and minimize random error in field trials. A field's natural variation is a major contributor to random errors in such trials and has been documented and worked upon in long-term trials in Rothamstead from the early 19th century. The early response to counter these variations was to create a specific plot size and "scatter" the plots, as suggested by Mercer and Hall [1]. Presently, spatial randomization is frequently employed to account for these natural variations. It has been the practice to account for the field's natural variability since its detailed intro- duction by Fisher [2], [3]. It was even more popularized by acclaimed experimental work from Rothamstead trials and is documented by Yates in [4]. Fisher's randomization used Latin squares to lay its plots in two directions of variability during the randomization of field trials. However, he had set the stage for blocking in randomized designs. Currently, randomized complete block design (RCBD) is extensively used to randomize field trials by using topographic variations as blocks, described in studies conducted by Borges et al. [5], Hoefler et al. [6], Casler et al. [7], Van Es et al. [8]. Various other studies have explored spatial blocks relying on natural variation indicators.

For example, Khosla [9] suggests soil electrical conductivity zones in central Pivot farming, similarly other soil properties and yield were used in studies by Borges et al. [5], King et al. [10], Galambošová et al. [11], Layton et al. [12], Cox et al. [13], Ortega et al. [14], Fleming et al. [15], Taylor et al. [16], Scudiero et al. [17] to account for a field's natural variation as blocks. However, yield or soil only partially represents the field's natural variations. A topographic (terrain and soil) view of the field can be a better measure of its natural variability. Authors such as Zhu et al. [18], Huang et al. [19], Khakural et al. [20], and Leuthold et al. [21] shared similar sentiments in their work. This creates a pressing need to design geospatial blocks based on local topographic features to account for the bias due to natural field variability in yield. Many studies have explored the need to include the effects of topographic variables such as soil and terrain on yield trends at field scale and found a significant correlation between yield and slope, elevation, Topographical Position Index (TPI), Topographic Wetness Index (TWI), soil texture, cation exchange capacity (CEC), bulk density, etc. Authors Leuthold et al. [21], [22], Peukert et al. [23], Jiang et al. [24], Guo et al. [25], Kravchenko et al. [26], [27] and Rahmani et al. [28] have explained in details the advantages of including composite topographic features in quantifying fertility. The yield outcome can be described primarily as a function of the amount of water and solar radiation available to the plant during its various growth stages. The solar radiation and other weather variables for field-grown crops are independent of human control and act like a nuisance factor. However, recent technological advancements have eased the use of topographic parameters at the field scale. Building on these, a system to analyze the field scale effect of topographic features on yield is essential. To this end, variations in topographic features such as elevation, slope, convexity, TPI, soil texture, drainage class, parent material, etc. were explored at the field level. This was followed by analyzing the individual effects of selected topographic features on yield for multiple years. This chapter describes the initial data exploration and descriptive analysis results of possible various topographic factors on yield. In this chapter, the section 1.2 describes the field used in this study, section 1.3 describes the preprocessing of topographic data, section 1.5 describes the analysis of topographic features in field 57 as a case study, and section 1.6 details the results of analyzing topographic effects on yield.

## 2    Study Area and Data Collection

This study was conducted at Purdue University's Agronomy Center for Research and Education (ACRE), encompassing an agricultural field of 3.5km × 3.5km. This farm consists of multiple small fields used by the University for agricultural field trials. ACRE's DEM was obtained from USGS digital elevation map (DEM) Light Detection and Ranging (LiDAR) database hosted on The National Map (NMAP) website. The NMAP provides elevation data with a spatial resolution of 1m. The LiDAR data for this dataset was collected in 2012.The vertical accuracy of the DEM data was 10cm or 4 inches. Soil classification data were sourced from the SSURGO soil database hosted by the USDA-NRCS and SoilGrids database hosted by the International Soil Reference and Information Centre (ISRIC). ISRIC SoilGrids was used to request higher-resolution soil data when available. The SSURGO soil database has a resolution of 1: 20, 000 (1 centimeter = 0.2 kilometer) scale on a 7.5- minute quadrangle US base map and SoilGrids has 100m spatial resolution available at selected location. Soil data about land classification, soil health, vegetative productivity, and land management were selected for this study. Inferences and soil characteristics such as soil texture, drainage class, component name, soil order, taxonomy class, taxonomy suborder, runoff, texture, crop productivity indices, moisture and temperature regimes, and parent material were important from the perspective of information required for classifying soil heterogeneity. Further processing of this data was required before using it in the project. These processing methods are described in Section 1.3 and Section 1.4. The historical yield was collected using a combine harvester during regular harvesting operations in 2015, 2017, and 2021

## 3    Data Preprocessing

The LiDAR-derived 1m resolution DEM data from NMAP was downloaded as tiles of raster images in the Tagged Image File Format (TIFF). The complete DEM of ACRE spanned across multiple tiles. The downloaded tiles were merged to create one raster file. This merged file was reprojected to the Universal Transverse Mercator (UTM) system of coordinates of Indiana zone 16. The Soil SSURGO data was retrieved using the soil data access (SDA) API of SSURGO. The coordinate reference system of SSURGO by default uses the Albers Equal Area projection centered on the Continental United States, based on the North American Datum of 1983 (NAD83). This was also reprojected to the UTM coordinate system of Indiana's zone 16. All preprocessing was done in Python.

## 4 High-resolution geo-indexed elevation and Soil data

The agricultural field was represented as a high-resolution 2-dimensional grid, where each grid cell corresponds to a specific area of land measuring 1m × 1m. The high-resolution elevation data were sampled to lower resolutions of (5m x 5m) and (10m x 10m). The lower-resolution data were used to analyze the effect of resolution on spatial analysis and infer the optimum sampling resolution. Each grid is characterized by a vector of terrain features. These are gradients and aspects derived from the elevation data. The elevation distribution of ACRE is illustrated by mapping the terrain colors on grayscale. The range of grayscale was mapped to the range of elevation data using histogram bins. The elevation data range was divided into 20 bins. The higher number of bins facilitates visualizing the smaller changes in elevation with greater effect. A smaller number of bins can mask the change of elevation, especially in flatter regions. The negative gradient of the elevation is illustrated in Figure 1(a) using black arrows. The length of the arrow represents the magnitude of the gradient, and it points to the direction of the slope. The top of the figure is geographically north, and the arrow rotates clockwise to illustrate the slope direction (aspect). The slope and aspect values were calculated using the equations (1) and (3). Let's consider a 3 × 3 matrix A

$$A = \begin{bmatrix} e_{00} & e_{01} & e_{02} \\ e_{10} & e_{11} & e_{12} \\ e_{20} & e_{21} & e_{22} \end{bmatrix}$$

Here, $e$ = elevation in meter (m).

$$slope = \sqrt{\frac{\partial e^2}{\partial x} + \frac{\partial e^2}{\partial y}} \qquad (1)$$

the gradient was calculated for grid/element of the elevation matrix using the expression,

$$(\nabla A)_{ij} = \left( \frac{A_{i,j+1} - A_{i,j-1}}{2\Delta x}, \frac{A_{i+1,j} - A_{i-1,j}}{2\Delta y} \right)$$

Here, $\Delta x$ and $\Delta y$ are the horizontal and vertical distance between grids. The one-side difference (using the first and the last column) expression is used for the edge cases.

$$(\nabla A)_{i0} = \left( \frac{A_{i1} - A_{i0}}{\Delta x}, \frac{A_{i+1,0} - A_{i-1,0}}{2\Delta y} \right)$$

$$(\nabla A)_{i,n-1} = \left( \frac{A_{i,n-1} - A_{i,n-2}}{\Delta x}, \frac{A_{i+1,n-1} - A_{i-1,n-1}}{2\Delta y} \right)$$

The Euclidean distance expression is used to calculate the final slope values shown in equation 2. The arc tangent of the vertical and horizontal gradients is used to calculate the aspect.

$$|\nabla A_{ij}| = \sqrt{\left(A_{i,j+1} - A_{i,j-1}\right)^2 + \left(A_{i+1,j} - A_{i-1,j}\right)^2} \quad (2)$$

$$aspect = \arctan \frac{-\frac{\partial e}{\partial y}}{\frac{\partial e}{\partial x}} \quad (3)$$

$$\theta_{ij} = \arctan\left(\frac{-\frac{A_{i+1,j} - A_{i-1,j}}{2\Delta y}}{\frac{A_{i,j+1} - A_{i,j-1}}{2\Delta x}}\right) \quad (4)$$

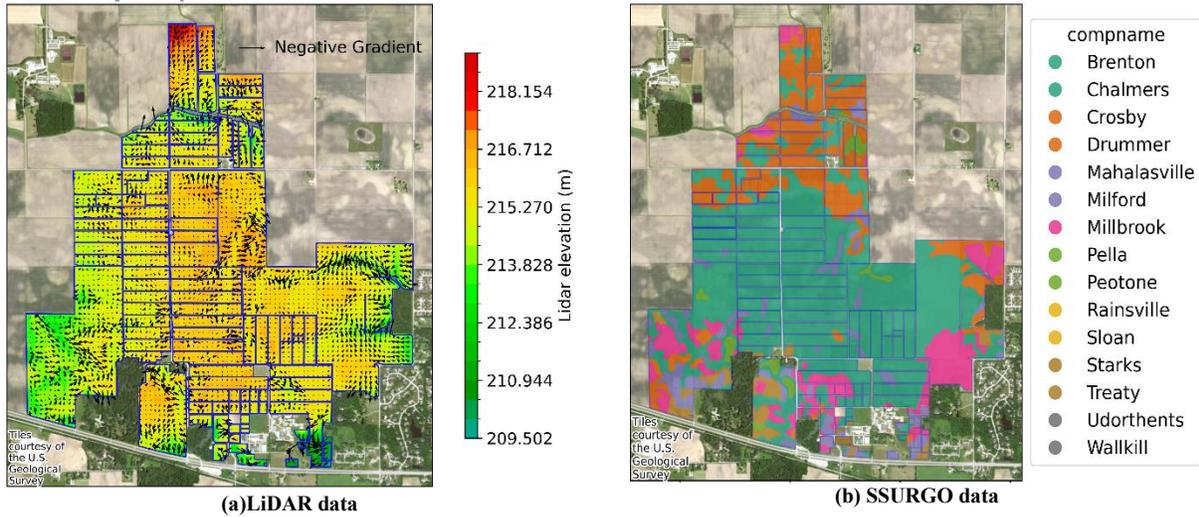

(a) LiDAR data  (b) SSURGO data

**Figure 1. Vectorized data from ACRE**

The grid vector of elevation and slope were merged with SSURGO soil data to create a comprehensive topographic dataset for ACRE. The SSURGO soil data described in Section II was preprocessed to merge the tabular data with the spatial data. The database structure of SSURGO [29] was used to select relevant tables for representing the soil's physical properties which influence its water-holding capacity. Data columns such as comp, mapunit, chorizon, chtexgrp, chtext, chtexmod, muaggatt, cpmat, cpmatgrp, soyNCCPI, cornNCCPI, sand content, silt content, clay content was retrieved using the soil data access (SDA) API wherever available. The SSURGO data were resampled and interlinked with the elevation point data at 1m, 5m, and 10 m resolution. The subfigures in Figure 2 show a few of the soil parameters mapped for ACRE. The subfigures used in this study were plotted from the gridded/vectorized dataset using Python programming language. Each element/row of the dataset represents a grid on the field. Every grid in the vectorized SSURGO dataset has columns defining several soils physical properties such as soil texture, drainage class, parent material, soil order, soil suborder, etc., soil chemical proper- ties such as k- saturation, electrical conductivity, cation exchange capacity (CEC), organic matter, etc., and soil productivity indices. Geographical manipulation techniques were used to sort the 3DEP data into SSURGO map unit polygons. Each row of the gridded terrain dataset was merged with the physical properties of the corresponding soil map unit from the merged soil dataset. These steps were implemented to create a high-resolution 2D gridded topographic vector dataset. These steps were repeated for the 5m × 5m and 10m × 10m resolution data. In the 2D gridded topographic vector dataset, the soil component name is the soil series name of the point. The most common soil series in ACRE is Brenton and Chalmers, see Figure 1(b) which belongs to the Mollisol soil order. In the soil order map, shown in Figure 2(a), Mollisols are the dominant soil order. Mollisols in Indiana are poorly drained soils specifically of the Aquolls suborder, see Figure 2(b). These soils have the seasonal high water table which keeps the groundwater near the topsoil level, resulting in wet soils in ACRE for most of the year, also described in the survey report of Tippecanoe soils by the soil survey staff [30]. The major drainage classes in ACRE are poorly drained shown in Figure 2(c). The soil texture is defined by the percentage of sand, silt, and clay in the soil pedon while sampling. The texture of the soil is also important in defining the water-holding capacity of the soil. Clayey soils hold more water than silty or sandy soils which can be good during low rainfall season, but they also appear to restrict plant germination due to water logging in average or good rain. The dominant soil textures in ACRE are shown in Figure 2(d).

# 5    Case study of Field 57 in Purdue ACRE Farms

Field 57 is a 625 m × 406 m field (Area= 253691.96 sqm, 62.7ac) on the east side of ACRE. Field 57 has been cultivated by rotating corn and soybeans since 2015. The soil and terrain data for field 57 was filtered from the ACRE raw data using geometric manipulations. The points of the merged High-resolution elevation data and resampled lower- resolution data (5m × 5m) and (10m × 10m) calculated for ACRE were filtered for field 57. The merged topographic dataset at three resolutions was created for Field 57 by selecting data from the gridded high resolution LiDAR data.

## 5.1    Topographic variation in Field 57

Satellite imagery of Field 57 from 2018 is shown in Figure 3(a) and 2024 is shown in Figure 3(b) and topographic features in Figure 3 (c) and 3(d). The distinct features in the north (A) and south- west parts (B) of the field appear to be lower elevation areas or regions with high soil moisture causing comparatively darker color in the satellite images. The elevation range in field 57 is ≈ 3m. The elevation in field 57 is plotted using 4 bins. The slope magnitude was represented by the length of the arrows and the aspect by the rotation of the arrow, see Figure 3 (c). The slope and aspect values were calculated using the 10m resolution grids in Figure 3 (c) for better visual representation. The A and B features from the satellite imagery were observed as low areas with higher values of negative slope in the Figure 3 (c). The soil components found in field 57 are shown in Figure 3 (d). The soil properties figures show the map units outlined with dashed lines and properties with shaded areas. Out of the four components observed, Chalmers was the dominant soil series in field 57. There are approximately 20,000 official soil series (listed as soil component names in SSURGO) in the USA defined by the USDA-NRCS. The official soil series describes Chalmers soil with the taxonomic class "fine-silty, mixed, super- active, mesic Typic Endoaquolls" [31]. This suggests that the soil in the Chalmers polygon has predominantly fine-silty soil texture, with a mixed mineral class. A mixed mineral class suggests an absence of any dominant mineral. It is super-active in its cation exchange capacity and has a mesic soil temperature regime. The soil temperature ranges from 8°C to 15°C in the mesic temperature regime.

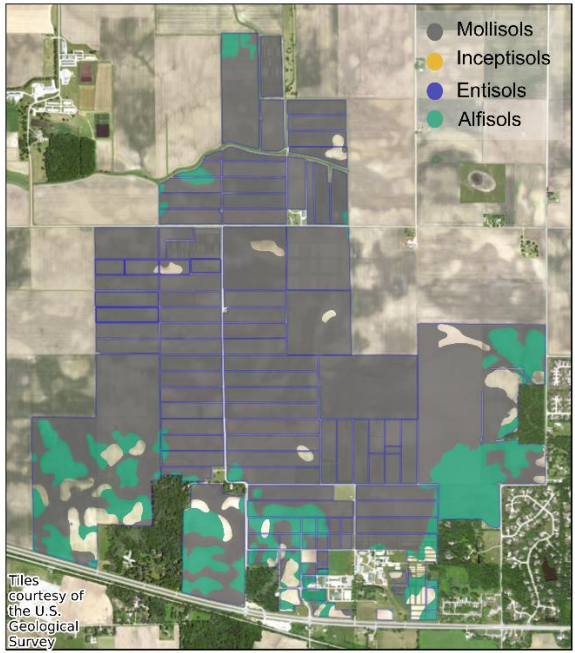
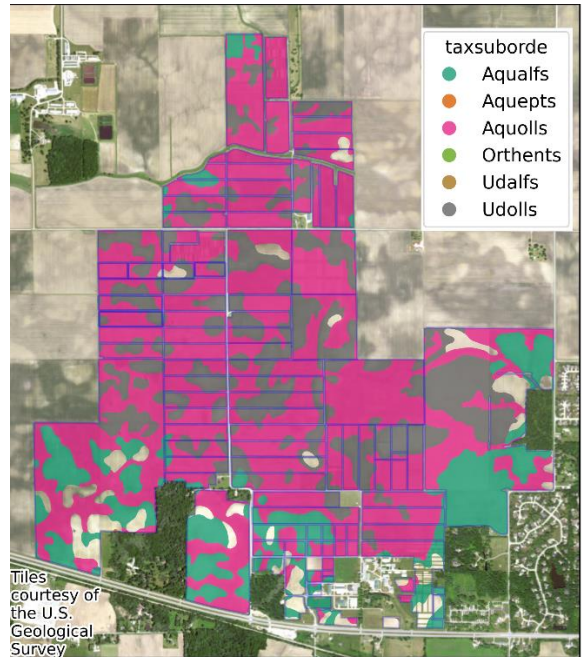
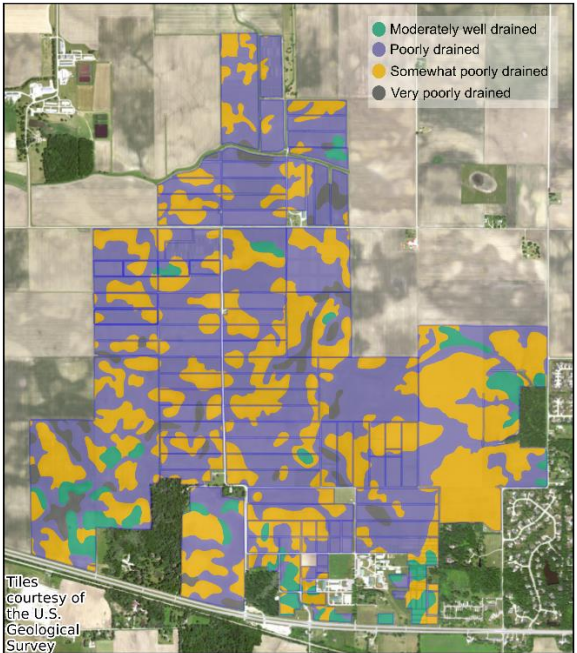
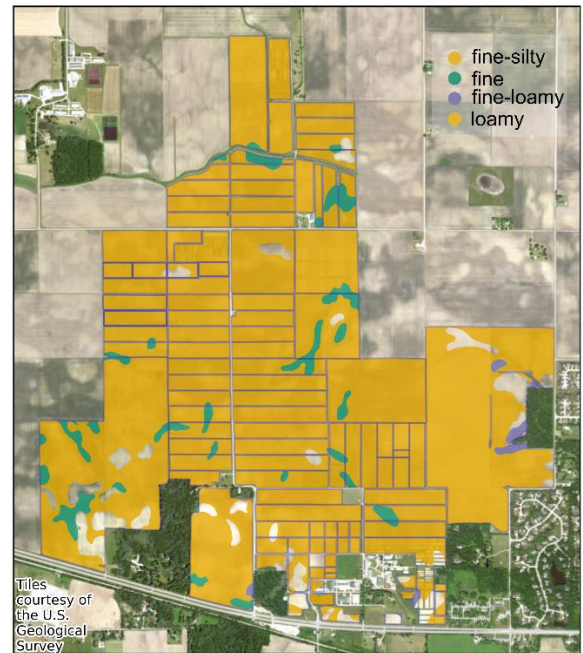

**Figure 2. (a) Tax-order (b) Tax- suborder (c) Drainage class (d) Particle size**

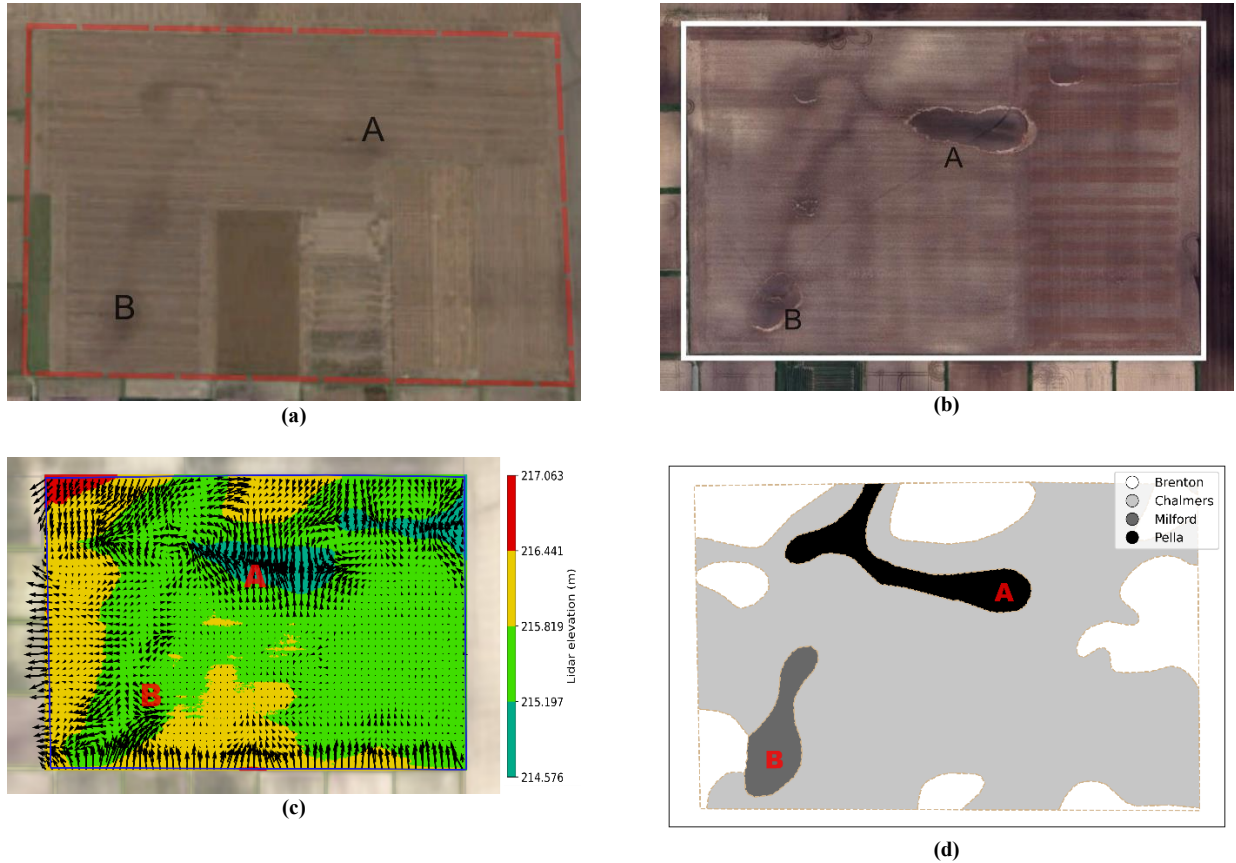

**Figure 3. Satellite images and topography of field 57.**

The overall character of the Chalmers is of a typical aquic Mollisol. These have high water tables throughout the year and are made up of thick dark topsoil high in organic mate rial. Prairies are the native vegetation of Chalmers soils. When drained artificially it supports mainly corn and soybean crops. Mollisols, the soil order of Chalmers, in Indiana, are remnants of glaciated lands of the north and are maintained with artificial drainage when used as croplands. The soil component in feature A is Pella [32] and B is Milford [33]. Pella is from the same taxonomic class as Chalmers, but unlike Chalmers, it is formed on depressional areas of ancient lake plains, outwash plains, and till plains. They also have carbonates in the top 102 cm of topsoil which is absent in Chalmers. The soil component in B is Milford of "Fine, mixed, super active, mesic Typic Endoaquolls" taxonomic class. These are formed on low broad summits and depressions in ancient glacial lakes. Milford has a higher clay content than both Pella and Chalmers. Brenton [34] soil belongs to the "Fine-silty, mixed, super active, mesic Aquic Argiudolls" taxonomic class. These are formed on outwash plains and stream terraces where the relief is relatively smooth. The overall clay content is less than Pella's and the same as Milford's. There are three drainage classes poorly drained, somewhat poorly drained, and very poorly drained in field 57 see Figure 4(a). Figure 4(c) shows the three parent materials in Field 57. Loess which is formed by silt depositions by wind are in the comparatively elevated regions. There are two major soil texture groups, silty clay loam and silt loam in field 57, see Figure 4 (b). Lacustrine deposits which represent low elevation areas in ancient lakes and cause clayey and poorly drained soils are observed in some parts of field 57. Most of the field 57 is formed over glacial till and Loess parent material.

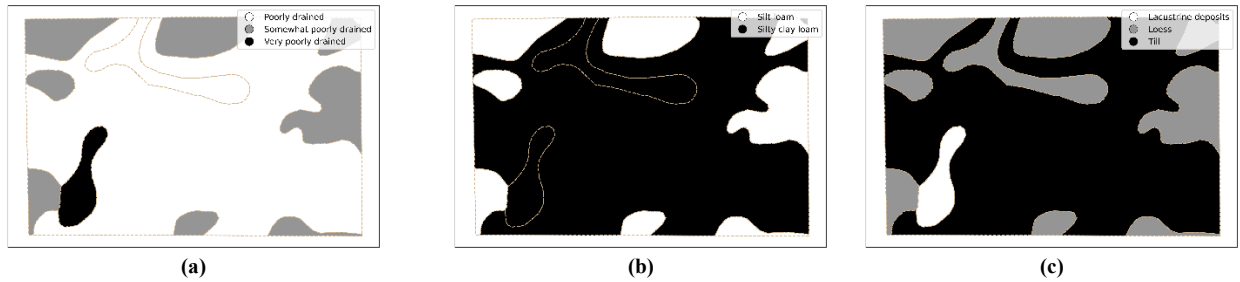

(a)  (b)  (c)

Figure 4. (a) Drainage class, (b) Soil texture, and (c) Parent material in field 57.

## 5.2 Data Contextualization

The soybean yield used in the topographical analysis requires us to add context to the factors affecting the yield in field 57. Soybean is a legume that fixes nitrogen from the atmosphere. This makes it suitable to rotate with corn. The best time to plant soybeans is from mid-April to early May. Soybeans grow best in warmer temperatures and well-drained loamy soils. Rain during the R1 to R6 reproductive stages of soybeans are suitable for a good yield. In Indiana weather, a soybean crop planted during late April and early May will reach its reproductive R1 stage during the third week of June and its reproductive R6 stage by mid-August. These times are crucial for the soybean crop. The average weather conditions in Indiana lead the average yield of soybean crops to be 57 bu/ac, one of the highest in the country. The yield data for the soybean crop in this field were obtained from the ACRE management in comma separated values (CSV) file format. The yield data was interpolated from actual yield data using the griddata method in Python. It is a spatial interpolation method which employs the Delaunay's triangulation and barycentric coordinates for linear interpolation. The interpolated values are the weighted average of the three nearest values. The weights reflect the distance of the values. The role of the Delaunay's triangulation is to ensure as equiangular triangles as possible. Given the actual yield data points at latitude and longitude values ex- pressed as $(x_i, y_i, yield_i)$ where $i = 1, 2, 3, \ldots, n$. We aim to interpolate the yield values onto new point $(x', y')$. The interpolated yield value at point $(x', y')$ is expressed by,

$$yield' = \lambda_1 yield_1 + \lambda_2 yield_2 + \lambda_3 yield_3$$

Where,
$yield'$ = interpolated yield at point $(x', y')$
$\lambda_1, \lambda_2, \lambda_3$ = Barycentric coordinates of point $(x', y')$, see Figure 5,
$yield_1, yield_2, yield_3$ = known values of vertices of the triangle.
From these the barycentric coordinates are calculated using the expressions

$$\lambda_1 = \frac{(y_2 - y_3)(x' - x_3) + (x_3 - x_2)(y' - y_3)}{(y_2 - y_3)(x_1 - x_3) + (x_3 - x_2)(y_1 - y_3)},$$

$$\lambda_2 = \frac{(y_3 - y_1)(x' - x_3) + (x_1 - x_3)(y' - y_3)}{(y_2 - y_3)(x_1 - x_3) + (x_3 - x_2)(y_1 - y_3)},$$

$$\lambda_3 = 1 - \lambda_1 - \lambda_2$$

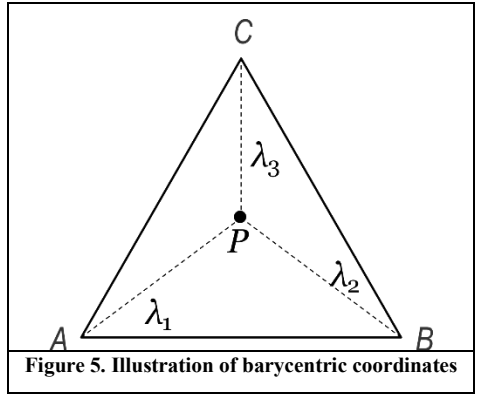

**Figure 5. Illustration of barycentric coordinates**

The yield in field 57 was interpolated to match the points in the gridded data. The interpolated yield data was spatially merged with the gridded terrain and soil characteristics. Thus, the field 57 has gridded topographic data for 2015, 2017, and 2021. Each row vector in this dataset now contains selected terrain information, soil physical properties, and yield data. The yield in Field 57 for three different years 2015, 2017, and 2021 was compared using Kernel density estimation (KDE) plots shown in Fig. 1.8. The kernel density estimation method was used to estimate the yield distribution. The Gaussian function was used in the smoothing kernel. The KDE uses the Gaussian function to calculate the distribution of the kernel centered at each sample of the data. Finally, the average of these distributions is the probability density function of the data, see equation 5.

$$f(x) = \frac{1}{\sigma\sqrt{2\pi}} e^{-\frac{(x-\mu)^2}{2\sigma^2}} \qquad (5)$$

Here,
$f(x)$ is the Gaussian function, μ is the mean of the distribution, and σ is the standard deviation. While implementing the KDE, the smoothing factor called bandwidth is incorporated in the Gaussian function. The bandwidth parameter is used to adjust the width of the kernel. The kernel width is important in representing the peaks in data optimally. Low bandwidths increase the spikes in the estimation resulting in a noisy density function whereas a large bandwidth value smooths the density function resulting in loss of information. To account for the bandwidth effect, a modified Gaussian kernel was used, see equation 6.

$$K_h(x - x_i) = \frac{1}{h\sqrt{2\pi}} e^{-\frac{(x-x_i)^2}{2h^2}} \qquad (6)$$

Silverman's thumb rule was used to calculate the bandwidth (h) for the Gaussian kernel. Silverman's factor for Gaussian kernel is given by equation 7 where, $h$ is the width of the kernel.

$$h = \left(\frac{4\hat{\sigma}^5}{3n}\right)^{\frac{1}{5}} \qquad (7)$$

Here, $\hat{\sigma}$ is the standard deviation of the data and $n$ is the number of samples. The bandwidth calculated from Silverman's rule of thumb, see fig. 1.8(a) were in the order of $10^{-2}$ whereas, bandwidth using std dev was in the order of $10^{1}$. This smaller bandwidth was able to display variations smoothed by the larger bandwidth. Silverman's factor is used in the KDE plots in the next sections. We observe the yield in 2015 is low but has a higher variation than other years. The rug plot at the bottom of the figures shows the density of the values on a linear scale. The weather is an important cause of yield variation. It varies the sunlight and moisture required for the growth of soybeans.

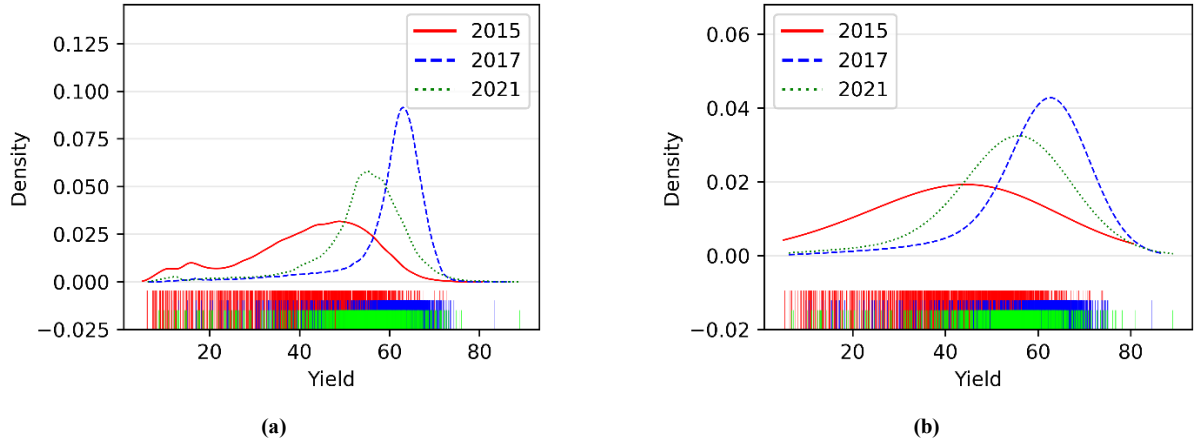

Figure 6. Yield variation in 2015,2017 and 2021 using (a)KDE plot with Silverman's bandwidth and (b) KDE with std dev bandwidth.

Several weather data models are used currently to represent the weather in high resolution using data from weather stations spread across Indiana state. Weather data was obtained from ACRE using public weather data processed by the Useful-2-Usable tool [35]. Even though weather is an important factor, controlling, measuring, or predicting its effect on yield is tedious and takes several years. For these reasons, we have used the weather as a nuisance factor in our two-way comparison of topographic effects on soybean yield in field 57. Nuisance factors are variables that affect the experiment but are not evaluated. The dark line represents the daily average temperature plot in 2015, see Figure 7(a), in 2017 see Figure 8(a), and in 2021 see Figure 9(a). These plots also show the variations in average daily temperature since 1984 using the shaded region. The average rain in ACRE is shown similarly in Figure 7(b), Figure 8(b) and Figure 9(b). The dashed vertical lines in the rain charts mark the usual soybean season in Indiana. The yield plots in Figure 7(c), Figure 8(c) and Figure 9(c) show the interpolated high-resolution gridded yield data from 2015, 2017, and 2021. The average yield in 2017 was 60 bu/ac. It is the highest among the three years followed by the average yield in 2021 with 54 bu/ac and 2015 with 42 bu/ac.

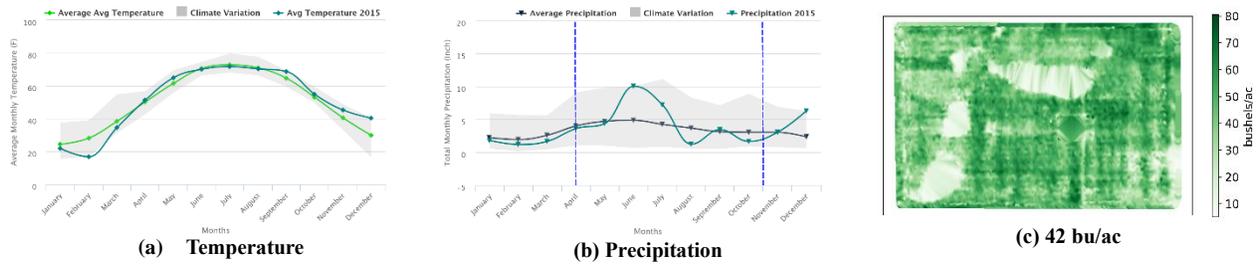

(a) Temperature  (b) Precipitation  (c) 42 bu/ac

Figure 7. Weather and average yield of field 57 in 2015.

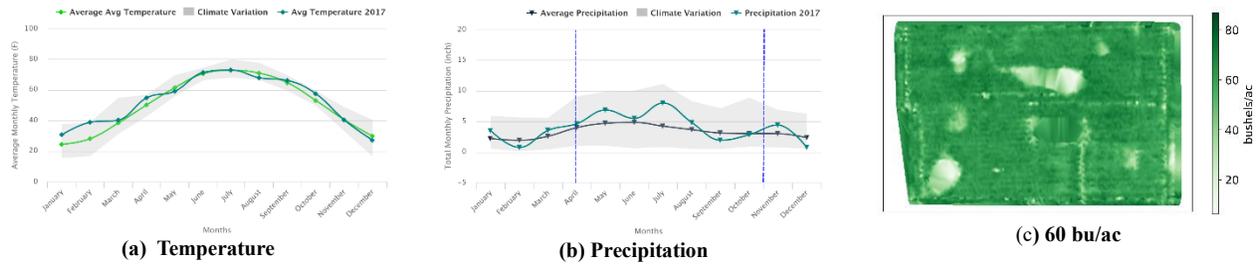

(a) Temperature  (b) Precipitation  (c) 60 bu/ac

Figure 8. Weather and average yield of field 57 in 2017.

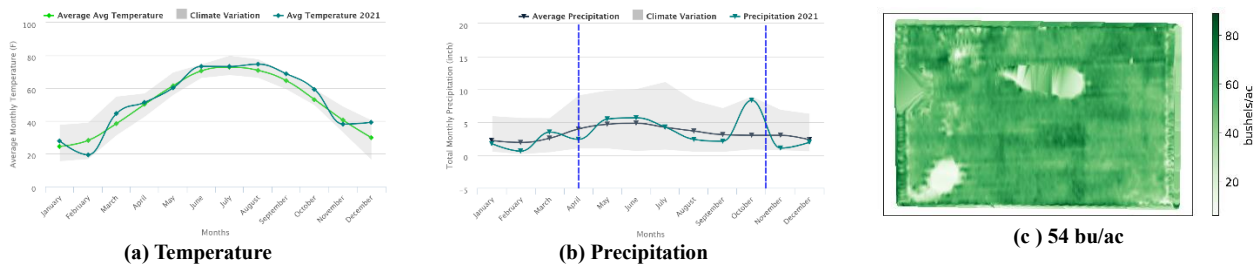

(a) Temperature  (b) Precipitation  (c) 54 bu/ac

Figure 9. Weather and average yield of field 57 in 2021.

The average temperature in 2015 is more than the historical average in September to October which is after the reproductive stage of soybean. The rain in 2015 was significantly more than the historical average rain in May, June, and July. It dropped to below average in the rest of the soybean season. 2015 was a low-yielding season. The temperature in 2017 was close to the historical average throughout the soybean season. The temperature was around 55°F during the planting weeks and increased steadily to 75°F. The temperature lowered from the average by a couple °F in mid-May and mid-August due to increased rain events. Rainfall in late May to early June was around 6 inches (≈ 2 inches more than average) and late July to early August was 8 inches (≈ 5 inches more than average). The temperature again increases from mid-September till the end of the soybean season with less than average (2 inches) rain, which is a good condition for drying soybeans. The weather conditions in 2017 appear to be favorable for soybean yield in Indiana. 2017 was one of Indiana's good soybean-yielding years with an average of 60 bu/ac. The average temperature and rain in 2021 followed a similar trend as 2017 but with less variation from the average. 2021 had an 8 inches rainfall which is ≈ 6 inches higher than average rainfall in the harvest season. More rain during the drying or harvest season increases the moisture in soybean yield lowering the harvest. The soybean in 2021 was less than in 2017 at 54 bu/ac. As we see here the weather conditions seem to explain the variation in the average yield of field 57, over three years, however on close inspection of the yield from field 57 we observe low-yielding areas in the field which seem to vary from 2015 to 2017 and 2021. Two or three areas on the field that coincide with lower elevation, high slope, and poor drainage seem to be performing worse than other areas irrespective of the weather. For example, 2015 had ≈ 5 inches more rain than average as well as more than in 2017 from mid-May till July. Yet, its average yield was lower than in 2017. The low yield may be attributed to low emergence in field 57's lower elevation- high slope-poor drainage areas due to more rain early in the soybean season. We have considered the effects of weather in inferring the effects of topography on yield in section 5.3.

## 5.3 Topographic effect on soybean yield in field 57

Land acts as the medium of water intake by the field crops. The topography of the land determines the water reaching plants by controlling the flow of water after a rain/irrigation event. Thus its effects are localized and vary frequently within season. On the other hand, the soil type determines its water-holding capacity. Soil types are determined by soil-forming factors and usually change naturally over centuries. These factors together regulate the major portion of water in the soil.

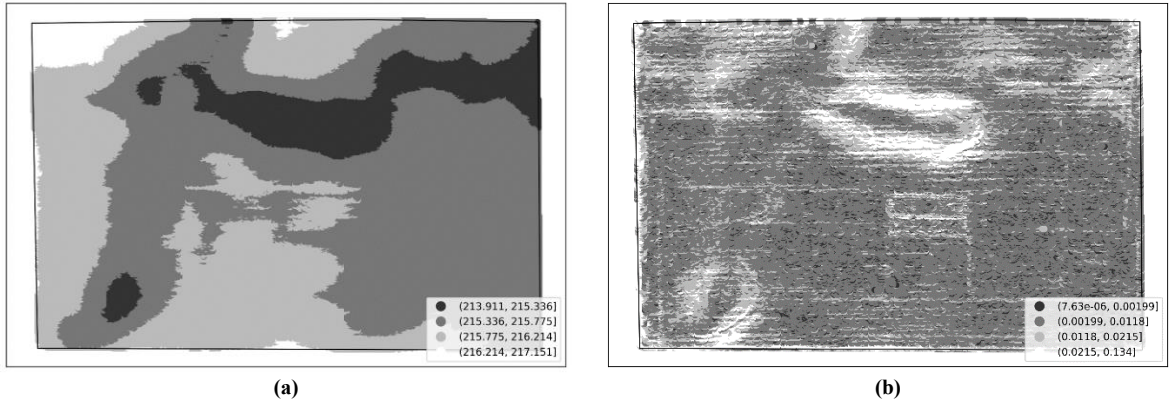

**Figure 10. (a)Elevation and (b) Slope groups in field 57**

The merged 2D gridded yield, terrain, and soil characteristics data were used to analyze the out- put yield in each unique topography (soil and terrain) character. Fig. 1.12 shows the elevation in Field 57 categorized into 4 continuous groups. The central tendency parameters of the elevation data [min, mean – std dev, mean + std dev, max] were used to group the elevation and slope value. Figure 10 shows two possible low areas in the field. It is interesting to observe here that the elevation groups match the soil polygons from the SSURGO dataset. The variation in the elevation in this field is ≈ 3 m. The box plot in Figure 11 represents the yield data distribution from the 4 elevation groups in field 57. In the box plot the lower quantile (25%) of the yield data is represented using the lower edge of the box and the upper quantile values are represented using the upper edge of the box. The box length is represented by the interquartile (IQR) range of the data, which is the middle 50% of the yield data. The line in the box represents the median of the yield data. The triangle marker represents the average yield value. The whiskers at the end of the line extending from the boxes are drawn at $1.5 \times IQR$. The whiskers in the box plot represent the extreme values of the typical distribution of the yield data. The upper whisker represents the larger of the max value or the $75^{th}$ percentile $+ 1.5 \times IQR$ of the yield data. Similarly, the lower whisker represents the smaller of the $25^{th}$ percentile $- 1.5 \times IQR$ or the minimum value of the typical yield data. The data points higher or lower than the typical distribution is considered outliers and represented by black rings. The solid trend line passing through the boxes joins the means of each yield data distribution. The dashed line with * represents a significant difference between the means of the groups.

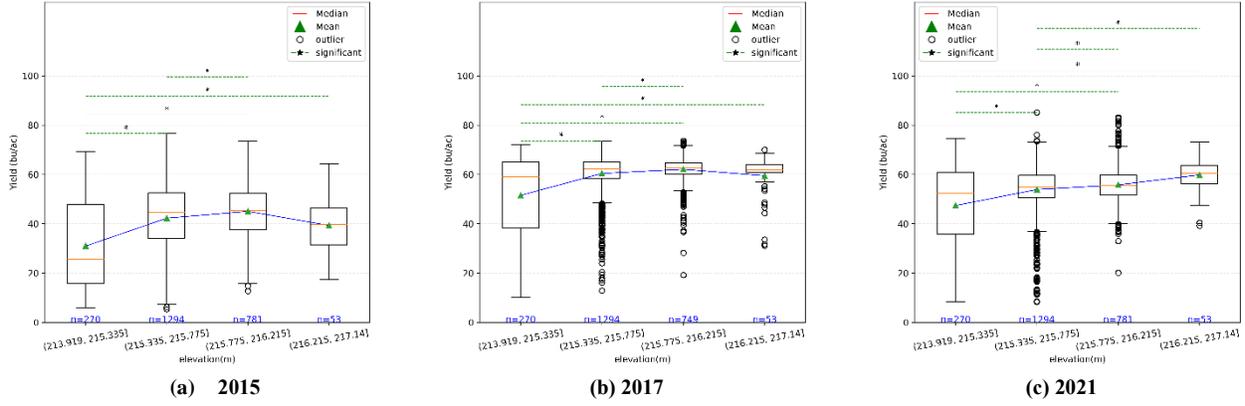

|   (a)   2015   |   (b)   2017   |   (c)   2021   |

**Figure 11. Yield box plots grouped by elevation classes in field 57.**

The number of yield data points (number of samples) is annotated at the bottom of each box in the plot. From the box-whisker plot, we observe that the yield was high in areas of field 57 with higher elevation and appeared to follow the expected trend of increasing yield with increasing elevation in 2021, see Figure 11(c). Whereas a definite decrease in average yield in the highest elevation region is shown by a dip of the mean-trendline in the last box of the plots in 2015 Figure 11(a) and 2017 Figure 11(b). The elevation group (215.335, 215.775] covers the largest area and has the highest number of yield data samples. It still shows consolidation around the mean group yield in 2017 and 2021 albeit with outliers. The lower areas of field 57 (213.919,215.335], display a high variation of yield in the analyzed three years. The yield variation is high in all elevation groups in 2015, a low-yielding year. The significance lines added to the box plot are described in detail in the section 5.4.

### 5.4  ANOVA analysis with Tukey's test for significance

#### 5.4.1  ANOVA of yield grouped by Elevation

Analysis of Variance (ANOVA)'s two-way analysis was used to test the null hypothesis that topographic variables do not affect yield. Ordinary Least Squares (OLS) regression was utilized to model the relation between yield and elevation in 2015, 2017, and 2021. The model shows low indication of explaining yield variability based on elevation alone. This is understood from the low R- squared value of 0.091 in 2015, 0.114 in 2017, and 0.061 in 2021. These values indicate that the OLS regression model could explain the yield variation correctly with 9.1% accuracy in 2015, with 11.4% accuracy in 2017, and with 6.1% accuracy in 2021. These values do not fall in the acceptable range for creating a good fit regression line. However, the model parameters from the OLS regression suggested a significant effect of elevation groups on yield in 2015, 2017, and 2021. The OLS model parameters can be used with ANOVA test results to confirm significant group differences. High F-statistics value and low probability of the F-statistics along with p-values below α = 0.01 are good indicators of significance in group differences. The expressions used in calculating the F-statistics for the independent variable, elevation groups, and dependent variable, yield. Sum of Squares Between (SSB) elevation groups:

$$SSB = \sum_{i=1}^{k} n_i (\bar{y}_i - \bar{y})^2 \qquad (8)$$

Here, $n_i$ is the number of observations in the $i-th$ group, $\bar{y}_i$ is the mean yield of the $i-th$ group, $\bar{y}$ is the overall mean of the yield, and $k$ is the number of groups ,4 in this case. Sum of squares of elevation groups is expressed with:

$$SSW = \sum_{i=1}^{k} \sum_{j=1}^{n_i} (y_{ij} - \bar{y}_i)^2 \tag{9}$$

Here, $y_{ij}$ is the yield of the $j-th$ observation in the $i-th$ group. From these the sum of squares total (SST) is computed by, $SST = SSW + SSB$, mathematically it is expressed as,

$$SST = \sum_{i=1}^{k} \sum_{j=1}^{n_i} (y_{ij} - \bar{y})^2 \tag{10}$$

The Degrees of Freedom between groups of elevation is 3; within the groups $(270 + 1294 + 781 + 53) - 4 = 2394$; and for the model is 3.

$$df_{between} = k - 1, df_{within} = N - k, \text{ and } df_{total} = N - 1$$

The Mean Square between groups (MSB) and Mean Square Within groups (MSW) are calculated with

$$MSB = \frac{SSB}{df_{between}}, \quad MSW = \frac{SSW}{df_{within}}$$

The F-statistics of the model is finally written as,

$$F = \frac{MSB}{MSW} \tag{11}$$

Tables 1, 2, and 3 show the test results of the two-way Analysis of Variance of yield in elevation groups. The high F-values along with the low probability of high F-values denoted by Pr(>F) indicate the significant difference in the mean yield of the elevation bin categories. The degree of freedom (*df*) is calculated for the model. The high values of the SSB, between categories (here elevation groups) in all three years are lower than the SSW, within the elevation groups (residual) re-indicating the model's inability to predict yield variations on elevation data alone. However, the lower SSB values indicate a better representation of yield variation by groups.

Table 1. Analysis of Variance of Yield grouped by elevation in 2015

| Source | Sum of Squares | $df$ | F-value | Pr(>F) |
|---|---|---|---|---|
| C(elevation bin) | $4.06 \times 10^4$ | 3.00 | 79.83 | < 0.001 |
| Residual | $4.06 \times 10^5$ | $2.39 \times 10^3$ | - | - |

Table 2. Analysis of Variance of yield grouped by elevation in 2017

| Source | Sum of Squares | $df$ | F-value | Pr(>F) |
|---|---|---|---|---|
| C(elevation bin) | $2.3 \times 10^4$ | 3.00 | 101.32 | < 0.001 |
| Residual | $1.78 \times 10^5$ | $2.36 \times 10^3$ | - | - |

Table 3. Analysis of Variance of yield grouped by elevation in 2021

| Source | Sum of Squares | df | F-value | Pr (>F) |
|---|---|---|---|---|
| C (elevation bin) | $1.60 \times 10^4$ | 3 | 51.88 | < 0.001 |
| Residual | $2.47 \times 10^5$ | $2.39 \times 10^3$ | - | - |

Analysis of Variance is followed by post hoc Tukey's honestly significant difference (Tukey's HSD) test to identify the groups with significant differences in average yield. Tukey's HSD method compares the mean yield of every possible pair of elevation groups. 4 groups of elevation can have 4C2 = 4!/(4 − 2)! × 2! = 6 combinations. The mean difference of each pair group was borrowed from the ANOVA table. The standard error of the mean difference is calculated from,

$$SE_{ij} = \sqrt{\frac{MSE}{n_i} + \frac{MSE}{n_j}}$$

here, MSE is the Mean Squared Error from ANOVA, and $n_i$ and $n_j$ are the number of observations in groups $i$ and $j$, respectively. The test statistic for the Tukey's HSD test is given by:

$$Q = \frac{\text{Mean Difference}}{SE}$$

The hypothesis is rejected if the Q value for the pairwise group is greater than the critical value from the studentized range distribution table at a 99% confidence interval or 0.01 significance level. The significance level $\alpha$ is replaced in Tukey's HSD by family-wise error rate ($FWER$). FWER accounts for the cumulative increase in the probability of Type I error in the multi-pair comparisons. This is countered by adjusting the significance level at each stage of comparison using equation 12, where $\bar{\alpha}$ is the adjusted p-value for pairwise comparison and $m$ is the number of independent comparisons. The adjusted p-value is deliberately kept conservative to limit within the overall FWER= 0.01. Another method of checking a true significance is, that the range of confidence interval of mean differences should not contain 0.

$$\bar{\alpha} = 1 - (1 - \alpha_{\text{per comparison}})^m \quad (12)$$

A mean difference equal to $0$ would only make the means equal and the null hypotheses would be correct. The 99\% confidence interval for the difference between the two means is:

$$\text{Mean Difference} \pm q_{\alpha,k,df} \cdot SE$$

here, $q_{a,k,df}$ is the critical value from the studentized range distribution table, at the significance level $\alpha$ of Tukey's HSD, number of groups, $k$, and degree of freedom, $df$ of the model. From these, the confidence interval is calculated as:

$$CI_{ij} = (\bar{x}_i - \bar{x}_j) \pm q_{\alpha,k,df} \times \sqrt{\frac{MSE}{n_i} + \frac{MSE}{n_j}} \quad (13)$$

here, $\bar{x}_i$ and $\bar{x}_j$ are the means of groups $i$ and $j$, and $n_i$ and $n_j$ are the sample sizes of the respective groups. Tables 4, 5, and 6 show the test results of post-hoc Tukey's honestly significant difference test performed pairwise on yield from elevation groups for 2015, 2017, and 2021. The tables show the difference in the average yield of each pair in the mean diff column. The adjusted p-value for $FWER = 0.01$ is shown in the p−adj column. The 99% CI column shows the confidence interval at the selected $FWER$ calculated using equation 13. The reject column lists the binary values for rejecting the Null hypothesis. The Null hypothesis can be rejected for the elevation group pair if its reject column value is True.

Table 4. Multiple comparisons of means of elevation groups with Tukey's HSD, FWER= 0.01 in 2015

| Group1 | Group2 | Mean Diff. | p-adj | 99% CI | Reject |
|---|---|---|---|---|---|
| (213.919, 215.335] | (215.335, 215.775] | 11.33 | 0.00 | [8.61, 14.04] | True |

| (213.919, 215.335] | (215.775, 216.215] | 14.09 | 0.00 | [11.23, 16.96] | True |
| (213.919, 215.335] | (216.215, 217.14]  | 8.38  | 0.00 | [2.28, 14.47]  | True |
| (215.335, 215.775] | (215.775, 216.215] | 2.76  | 0.00 | [0.93, 4.60]   | True |
| (215.335, 215.775] | (216.215, 217.14]  | -2.95 | 0.37 | [-8.64, 2.74]  | False |
| (215.775, 216.215] | (216.215, 217.14]  | -5.72 | 0.01 | [-11.48, 0.04] | False |

Table 5. Multiple comparisons of means of elevation groups with Tukey's HSD, FWER= 0.01 in 2017

| Group1 | Group2 | Mean Diff. | p-adj | 99% CI | Reject |
| --- | --- | --- | --- | --- | --- |
| (213.919, 215.335] | (215.335, 215.775] | 8.88  | 0.00 | [7.07, 10.70]  | True |
| (213.919, 215.335] | (215.775, 216.215] | 10.59 | 0.00 | [8.67, 12.51]  | True |
| (213.919, 215.335] | (216.215, 217.14]  | 7.95  | 0.00 | [3.88, 12.01]  | True |
| (215.335, 215.775] | (215.775, 216.215] | 1.71  | 0.00 | [0.46, 2.95]   | True |
| (215.335, 215.775] | (216.215, 217.14]  | -0.94 | 0.87 | [-4.73, 2.74]  | False |
| (215.775, 216.215] | (216.215, 217.14]  | -2.64 | 0.14 | [-6.49, 1.21]  | False |

Table 6. Multiple comparisons of means of elevation groups with Tukey's HSD, FWER= 0.01 in 2021

| Group1 | Group2 | Mean Diff. | p-adj | 99% CI | Reject |
| --- | --- | --- | --- | --- | --- |
| (213.919, 215.335] | (215.335, 215.775] | 6.49  | 0.00 | [4.38, 8.61]   | True |
| (213.919, 215.335] | (215.775, 216.215] | 8.38  | 0.00 | [6.15, 10.62]  | True |
| (213.919, 215.335] | (216.215, 217.14]  | 12.43 | 0.00 | [7.67, 17.18]  | True |
| (215.335, 215.775] | (215.775, 216.215] | 1.89  | 0.00 | [0.46, 3.32]   | True |
| (215.335, 215.775] | (216.215, 217.14]  | 5.93  | 0.00 | [1.50, 10.37]  | True |
| (215.775, 216.215] | (216.215, 217.14]  | 4.04  | 0.03 | [-0.45, 8.53]  | False |

No significant difference was observed in the average yield of the top two highest elevation groups (215.775, 216.215] and (216.215, 217.14] of field 57 in 2015, 2017, or 2021. The p-adjusted value did not satisfy the condition required for significance as they were not less than FWER in the pairwise comparison using Tukey's HSD. These pairs of elevation groups also fail the confidence interval test. It is interesting to note here, the results of the pairwise comparison of the third-highest region (215.335, 215.775] and the highest region (216.215, 217.14] of field 57 respectively. The pairwise comparison of the mean yield in these areas did not result in significant differences in 2015 and 2017 but the mean yield became significantly different in 2021. The difference in the mean yield between (215.335, 215.775] and (216.215, 217.14] in 2021 was greater than in 2015 or 2017. This pair also passed the significance test based on the confidence interval in 2021. The rest of the pairwise tests had a significant difference in the average yield and had a p-adjusted value was less than FWER in all three years. These also displayed significant differences in the confidence interval test. Notable observations while analyzing the effect of elevation on yield using descriptive analysis from box-and-whisker plots, analysis of variance in yield from four elevation groups, and post-hoc analysis results obtained using Tukey's HSD test on data from three years are noted below. The weather factors were included in the observations to draw plausible inferences.

–The high-elevation (216.215, 217.14] area, had a significantly better yield in 2021 observed from the box plot, Tukey's HSD test result, and fig. 3.11. 2021 had near-historical average early-season rain and high late-season rain. The elevated areas were able to drain the soil moisture from high late-season rain thus allowing soybeans to dry, for harvest. This area had a worse yield than the majority of the field (areas with elevation (215.335, 215.775] and (216.215, 217.14]) in 2015 and 2017, as seen in the box plots in fig. 3.13(a) and fig. 3.13(b) and was again inferred from the negative mean yield difference values in Table 3.4 and Table 3.5.

–The high SSB and SSW values from the ANOVA tables suggest that the yield variation can be explained better by the group effect of elevation rather than independently.

### 5.4.2  ANOVA of yield grouped by Slope

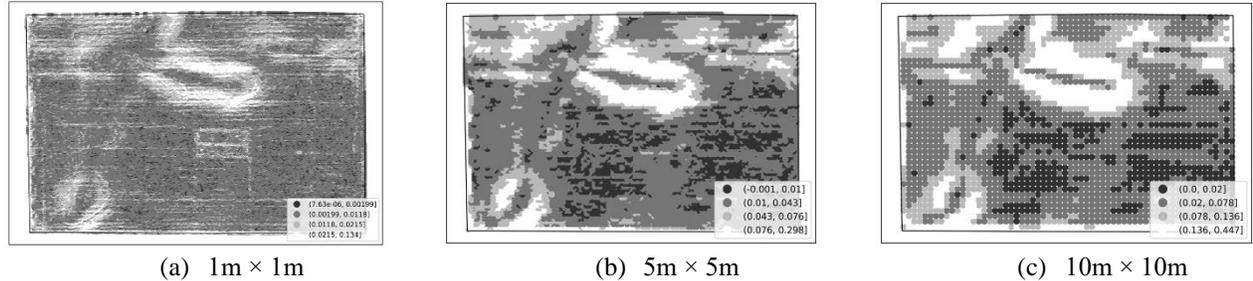

(a)  1m × 1m  (b)  5m × 5m  (c)  10m × 10m

**Fig. 3.14. Slope in field 57 calculated at three resolutions.**

The slope in field 57 was calculated from the gridded data at three resolutions. The gradients of these three resolutions were calculated from the elevation described in eq. (3.2) at three resolutions. The resulting slope at three resolutions was able to identify different topographic features in the field. The low elevation area in the field appears with comparatively higher gradient values depicted in lighter shades. Field 57 is mostly flat as seen by the extent of the low range of slope values (dark shade), see Fig. 3.14. While calculating the elevation derivatives, a challenge was observed. The 1m × 1m resolution slope identified small unwanted features, even the remnants of the crop rows made in the center of the field in previous years, see Fig. 3.14(a). However, the slope at this resolution could not classify the field into differentiable areas for group effect analysis. The same problem was observed in the slopes calculated at 5m × 5m resolution, but these were able to draw sharply defined edges around the low-elevation areas, see Fig. 3.14(b). This was important in deciding the optimum spatial resolution for calculating the slope. The KDE of slope values from the gridded topographic data of three resolutions were plotted, see Fig. 3.15. The slope values from 1m × 1m gridded data have minimum variation in the plot. The frequent spiky values from this plot detect minute slope changes which only make the data noisy. The noisy values here are not errors, instead they represent frequent changes in slope calculated from the high-resolution elevation data captured by high-end LiDAR sensors. The KDE plot for 5m × 5m has a much lower density and larger variation than the 1m resolution data. The 10m × 10m resolution gridded data creates a KDE plot with an even lower density value and large variation. The better grouping effect of the 10m × 10m resolution slope data and its less noisy character enables the identification of larger area effects.

This was also suitable for use with the available yield data, which was measured by yield monitor on combine harvesters at a similar resolution of ≈ 0.005ac [96]. The four slope groups in Field 57 were calculated using the same method as the elevation groups. In the box-and-whisker plot of the slope groups, the yield appears to follow the expected trend of decreasing with an increase in slope values.

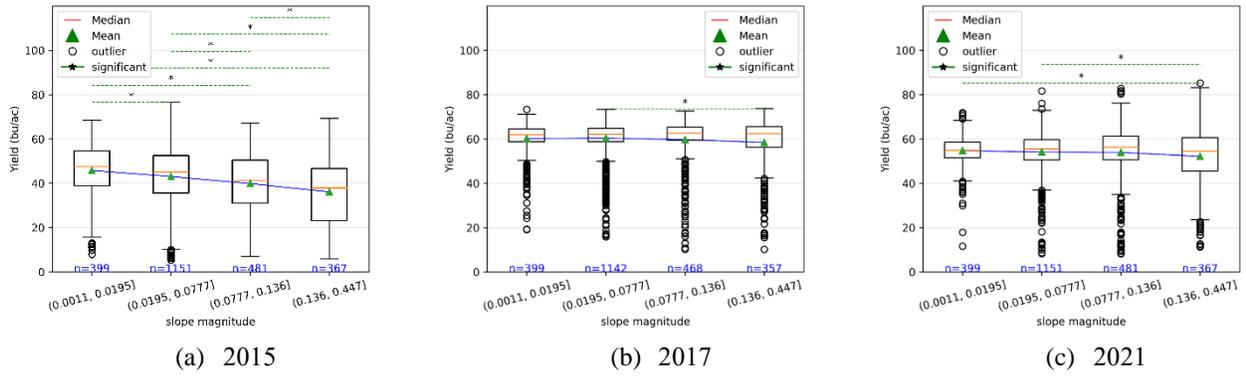

| | | |
|---|---|---|
| (a) 2015 | (b) 2017 | (c) 2021 |

**Fig. 3.16. Yield box plot grouped by slope classes in Field 57.**

This effect is most prominent in 2015, but not in 2017 and 2021. The yield in 2015 had a large variation in every slope range, however, the yield variations in 2017 and 2021 were small with many outliers. The results of the pairwise significance test identified a more significant group effect in 2015 as compared to 2017 and 2021. These are explained in more detail using ANOVA and post-hoc Tukey's HSD test. These tests were organized using the previously described methods from section 3.6.1 and were implemented on slope data. The test results are summarized in the tables below. The results of the Analysis of Variance of yield in four slope groups are displayed in table 3.7, table 3.8, and table 3.9.

Table 3.7. Analysis of Variance of yield grouped by slope in 2015

| Source | Sum of Squares | df | F-value | Pr(>F) |
|---|---|---|---|---|
| C (slope bin) | $2.19 \times 10^4$ | 3 | 41.13 | < 0.001 |
| Residual | $4.25 \times 10^5$ | $2.39 \times 10^3$ | | |

Table 3.8. Analysis of Variance of yield grouped by slope in 2017

| Source | Sum of Squares | df | F-value | Pr(>F) |
|---|---|---|---|---|
| C (slope bin) | $1.10 \times 10^3$ | 3 | 4.34 | - |
| Residual | $2 \times 10^3$ | $2.36 \times 10^3$ | - | - |

Table 3.9. Analysis of Variance of yield grouped by slope in 2021

| Source | Sum of Squares | df | F-value | Pr(>F) |
|---|---|---|---|---|
| C(slope bin) | $1.49 \times 10^3$ | 3 | 4.55 | < 0.001 |
| Residual | $2.61 \times 10^3$ | $2.39 \times 10^3$ | - | - |

The R − squared values of the OLS model of slope and yield were overall very low. This also suggests the model's inability to predict yield based on slope alone. In the ANOVA test results, the SSB in all three years are lower than the SSW. These values suggest that the higher variability of yield can be explained better between groups. The small difference between SSB and SSW also results in low F-statistics values for 2017 and 2021, however, the probability of the F-statistics values was still lower than the significance (α = 0.01) of these tests. The ANOVA test showed a significant effect of groups on yield in 2015, 2017, and 2021. The post-hoc test was required to identify the groups that had significantly different yields to complete this analysis. The post-hoc Tukey's HSD test compared the 6 pairs of 4 slope groups to identify the pairs with significant differences in their average yield. In 2015 all the groups had significantly different average yield however, this decreased to one pair in 2017 and two pairs in 2021. Notable observations while analyzing the effect of slope on yield using descriptive analysis from box-and-whisker plots, analysis of variance in yield from four elevation groups, and post-hoc analysis results obtained using Tukey's HSD test on data from three years are noted below. The weather factors were included in the observations to create plausible inferences.

–All the groups show significant differences in 2015. As we recall, 2015 had a high mid-season rainfall in field 57 and a low average yield of 42 bu/ac. The high variation in the yield from all slope groups indicates that the rain pattern of 2015 caused significant differences in the yield in slope groups. In comparison to the insignificant difference in the average yield between majority groups in 2017 and 2021 which had comparatively better rain patterns for soybean growth. This indicates that the effects of rain patterns are amplified on yield, based on slope groups.

–The slope alone cannot explain yield differences in field 57. The slope along with elevation can be used.

Having observed the effects of terrain on yield through slope and elevation analysis, it is crucial to examine the effect of soil. Soil serves as the medium for plant-water interaction and is equally important in the analysis of yield variations influenced by topography. For this study, key soil physical properties such as drainage class, soil texture, and parent material were selected which are vital in deciding the water available to plants. Soil's drainage classes are categorical variables. The SSURGO defines seven distinct soil drainage classes: Excessively drained, Somewhat excessively drained, Well drained, Moderately well-drained, Somewhat poorly drained, Poorly drained, and Very poorly drained. This classification reflects a gradient in soil drainage capability, decreasing from the highest in excessively drained soils to the lowest in very poorly drained soils. Soil texture classes are based on the ratio of sand, silt, and clay present in the soil. Soil scientists use the soil texture triangle to name the soil texture classes based on this ratio, see Fig. 1.17. Soil texture is essential to yield as it controls the inter-particle forces that control the movement of water in the soil. Plants use capillary action to absorb water from the soil while water moves around in the soil under gravitational force.

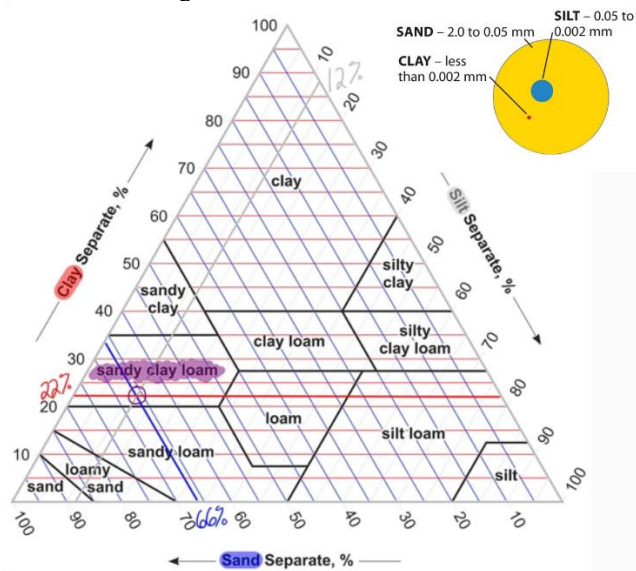

**Fig. 1.17. Soil Texture Triangle**

The inter-particle forces of soil particles counter these forces. These inter-particle forces are dependent on the ratio of air, water, and soil particles by volume. Sand particles are much larger than clay and silt particles and have large pore spaces. This allows large spaces for air and water in sandy soil. The large shape of sand particles creates less surface area in a unit volume, consequently weakening the inter-particle forces. These factors assist in draining water quickly and support aeration

On the other hand, clayey soils with smaller particle sizes are compactly packed in unit volume and have fewer water and air pockets. The small particle size increases particle density, which consequently increases the surface area and inter-particle forces. Hence, soils with more clay pro- portions are dense with very little or no aeration and high water holding capacity. Clays are made up of secondary minerals such as phyllosilicates, carbonates, sulfates, etc. which have complex bond structures. The complex structure and particle size of clay creates large inter-particle forces that resist the gravitational and capillary forces. This adds to the high water retention capacity and low aera- tion. These also affect the shrink and swell capacity of clays which further affects soil drainage. Silt particles are of intermediate size and behave accordingly. Thus the soil texture affects the yield in several indirect ways too. The parent material is the depth-limiting layer or the base layer of soil. Soils are formed either by weathering or deposition. Soil formation from parent materials is affected by climate, time, and re- lief. There are nine parent materials defined in the SSURGO data namely Alluvium, Colluvium, Eolian Sand, Lacustrine Deposits, Loess, Outwash, Residuum, Till, and Volcanic Ash. These will be described as and when necessary in this document. Soil parent materials are one of the five factors of soil formation. The parent material is a good indicator of the soil's physical properties, especially in cases of mixed soil use classes or where the topsoil has been disturbed. The soil physical characteristics data in field 57 obtained from SSURGO data lists three drainage classes see Fig. 1.18(a), two soil textures, see fig. 1.18(b) and three distinct parent material, see fig. 1.18(c).

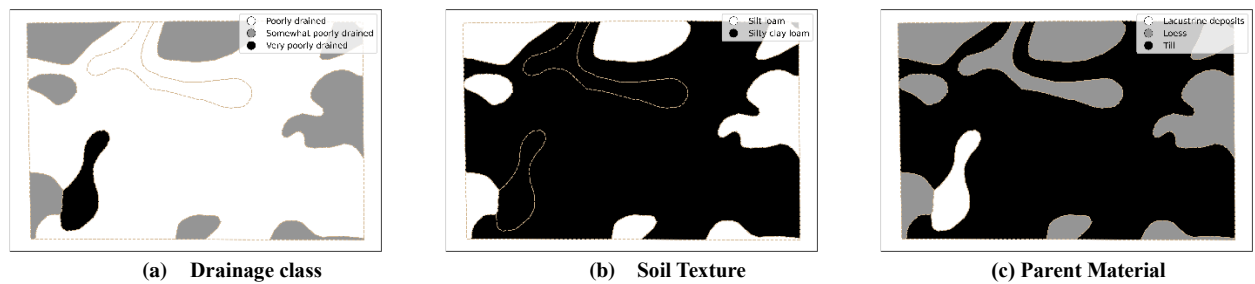

(a)  **Drainage class**  (b)  **Soil Texture**  (c)  **Parent Material**

**Fig 1.18. Soil physical properties of Field 57**

The polygons in the Fig. 1.18 show the different SSURGO map units in field 57 using dashed lines. The shaded regions illustrate the properties. The Somewhat poorly drained (SPD) coincides with the higher elevation areas in field 57 and silt-loam soil texture. Most of the SPD drainage is in soil made from loess parent material. Loess is silt deposited by wind. The Poorly drained (PD) class is extensively present in field 57 it coincides with the silty clay loam soil texture. The PD class of soil also extends over the flat slope areas and mid-range elevation. The dark-shaded region area (A) in the north, see fig. 1.5 is in the PD class even though it belongs to a different soil component because its parent material is Loess. Loess parent material is distinctly labeled only when at least $51 - 102$cm ($1.6 - 3.3$ft) thick silt is deposited by wind. Feature B in fig. 1.5 belongs to the silty clay loam texture. It is a somewhat low-lying area and has Lacustrine deposits of ancient lakes as its parent material. All these factors together make it a VPD soil. From these observations, we can infer that a soil's physical character heavily depends on its formation factors within a field also. Consequently, to analyze the effects of soil's physical properties on yield, we have employed the ANOVA and Tukey's HSD post-hoc analysis methods from section 1.6.1 on soil texture, drainage class, and parent materials.

*5.4.3  Yield analysis by Soil texture category*

The yield maps in figures 1.19(a), (b) and (c) use hatching to illustrate the yield in each soil texture in field 57. The box whisker plots in 1.19(a), (b), and (c) show the yield trend grouped by soil texture. The box- and-whisker plots show a higher soybean yield in silt-loam areas. These are also the areas that are PD and VPD and have low to medium elevation. The yield variation in 2015 was high for both groups of texture class. Following the trend of slope classes the yield variation in 2017 and 2021 shows a narrow variation albeit with many outliers. The texture class-interpolated yield OLS model showed low

R-squared values. This result aligns with the expectations set by previous variables in predicting yield using one of the topographic parameters alone. The ANOVA of the yield variation is summarized in table 1.13, table 1.14, and table 1.15. The F-statistics values are low yet its probability is less than α in 2015 and equal to α in 2017. These values are enough to indicate a significant difference in average yield caused by the texture class. The results in 2021 however, do not support the rejection of the null hypothesis. Tukey's HSD test results even though not required in this texture class analysis as there are only two texture categories, still let us understand the effect between group yield in three different years.

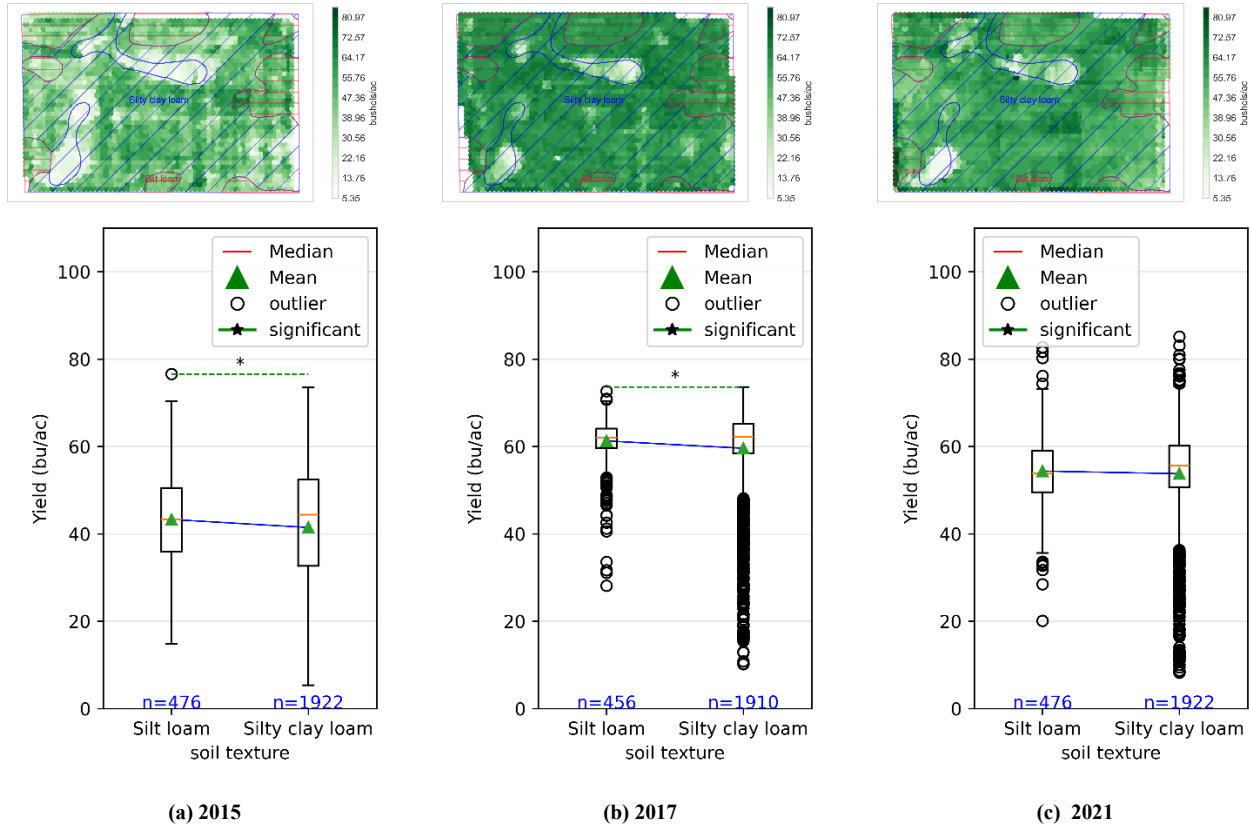

(a) 2015      (b) 2017      (c) 2021

Fig. 1.19. The yield is grouped by soil texture types in Field 57.

Table 1.13 Analysis of variance of yield grouped by texture description in 2015

| Source | Sum of Squares | df | F-value | Pr(>F) |
|---|---|---|---|---|
| C(texdesc) | $1.27 \times 10^3$ | 1 | 6.85 | 0.01 |
| Residual | $4.45 \times 10^5$ | $2.4 \times 10^3$ | - | |

Table 1.14 Analysis of variance of yield grouped by texture description in 2017

| Source | Sum of Squares | df | F-value | Pr(>F) |
|---|---|---|---|---|
| C(texdesc) | $1.05 \times 10^3$ | 1 | 12.43 | <0.001 |
| Residual | 2×10⁵ | 2.36×10³ | - | |

**Table 1.15 Analysis of variance of yield grouped by texture description in 2021**

| Source | Sum of Squares | df | F-value | Pr(>F) |
|---|---|---|---|---|
| C(texdesc) | 141.09 | 1 | 1.29 | 0.26 |
| Residual | 2.63×10⁵ | 2.4×10³ | - | |

**Table 1.16 Comparison of yield means from soil texture groups with Tukey's HSD FWER =0.01**

| year | Group1-Group 2 | Mean Diff | p-adj | 99%CI | Reject |
|---|---|---|---|---|---|
| 2015 | Silt loam-silty clay loam | -1.83 | 0.01 | [-3.63, -0.03] | Yes |
| 2017 | Silt loam-silty clay loam | -1.69 | 0.00 | [-2.93, -0.45] | Yes |
| 2021 | Silt loam-silty clay loam | -0.61 | 0.26 | [-1.99, 0.77] | No |

As discussed previously, soil phys- ical properties such as texture, drainage, etc. stay constant, however, their effects on yield are amplified in extreme weather patterns. The randomness of weather patterns makes it tricky to assume the effects of soil physical properties are a nuisance factor in designing field experiments. As the topographic effects on yield have not been standardized or measured for extreme weather patterns, a comparative analysis like Tukey's HSD paves the way for recording the soil's effect on yield even with limited time data. For example, Tukey's HSD test for soil texture categories suggests a higher sensitivity to weather patterns. This can be seen in the high p-adjusted value in 2021 and 2015. Both these years show deviation from the preferred weather pattern for soybeans seen in 2017. Though statistically the mean yield is significantly different in 2015, a low F-statistic value was observed in its ANOVA same as the test results of 2021. The low SSB values in 2021 (late-season rain) emphasize the inefficiency of a single topographic factor in explaining the weather-topographic effect on yield.

*5.4.4    Yield analysis by parent material category*

The Lacustrine deposit areas have a poor yield in all years Fig. 1.20. However, its effect is amplified with early rain as seen in the low mean yield in 2015, see fig. 1.20(d) and high mean difference in Table 1.22. The difference in average yield between lacustrine deposits and other parent material is high in all available years, see Fig. 1.20(e) and fig. 1.20(f). The till areas also outperform the loess areas. A comparatively higher mean difference was observed for till areas in Tukey's HSD test results, even though the loess areas are on comparatively higher elevations. The difference in mean yield from loess and till was lower in 2017 (had a favorable weather pattern). The analysis results of average yield by parent material summarize the significant effects of parent material on yield in field 57. The ANOVA and Tukey's HSD test results have suitable values to reject the null hypothesis, which does not change in the three available years of yield data from field 57. Thus, different parent material's influence on the yield outcome holds unless a more extreme weather pattern is observed. The OLS for the interpolated yield-parent material model had higher R-squared values- 14.3% in 2015, 11.6% in 2017, and 16.1% in 2021 as compared to previous variables, still these alone were not acceptable for predicting yield outcome behavior. The results of the ANOVA and post-hoc Tukey's HSD tests are summarized in Tables 1.14.

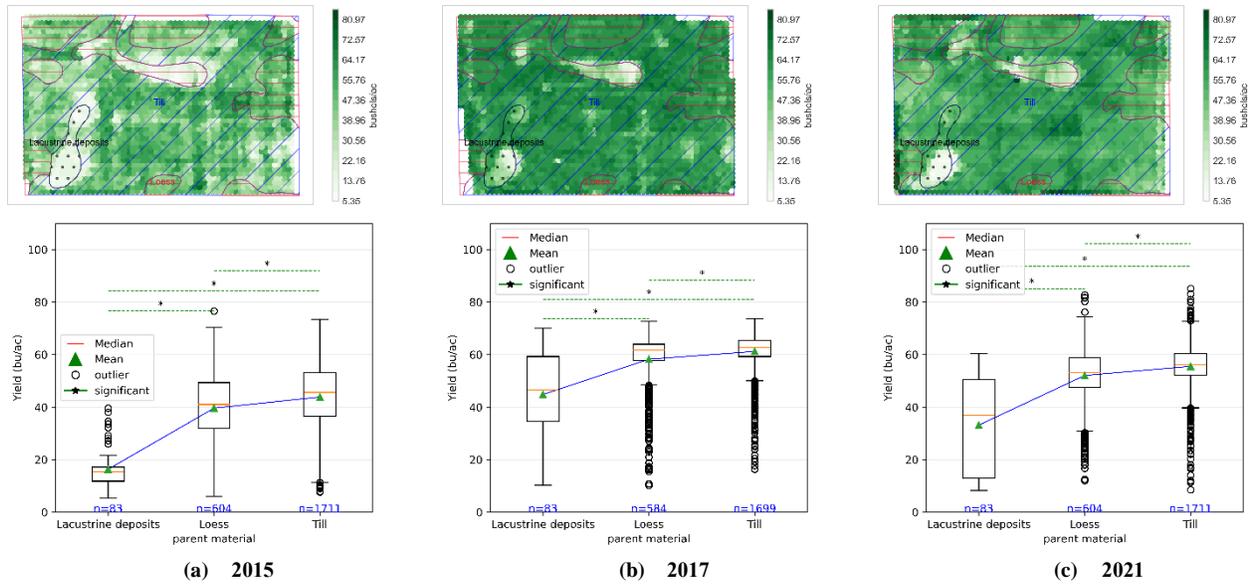

| (a) 2015 | (b) 2017 | (c) 2021 |

**Fig 1.20 Yield grouped by parent material in field 57**

**Table 1.17 Analysis of variance of yield grouped by parent material in 2015**

| Source | Sum of Squares | df | F-value | Pr(>F) |
|---|---|---|---|---|
| C(parent material) | $6.37 \times 10^4$ | 2 | 199.3 | <0.001 |
| Residual | $3.83 \times 10^5$ | $2.4 \times 10^3$ | - | - |

**Table 1.18 Analysis of variance of yield grouped by parent material in 2017**

| Source | Sum of Squares | df | F-value | Pr(>F) |
|---|---|---|---|---|
| C(parent material) | $2.34 \times 10^4$ | 2 | 155.59 | <0.001 |
| Residual | $1.78 \times 10^5$ | $2.36 \times 10^3$ | - | - |

**Table 1.19 Analysis of variance of yield grouped by parent material in 2021**

| Source | Sum of Squares | df | F-value | Pr(>F) |
|---|---|---|---|---|
| C(parent material) | $4.23 \times 10^3$ | 2 | 229.63 | <0.001 |
| Residual | $2.2 \times 10^5$ | $2.4 \times 10^3$ | - | - |

**Table 1.20 Multiple comparisons of yield mean from parent materials with Tukey's HSD, FWER= 0.01 in 2015**

| Group 1 | Group2 | Mean Diff. | p-adj | 99%CI | Reject |
|---|---|---|---|---|---|
| Lacustrine Deposit | Loess | 23.25 | 0.0 | [18.93, 27.57] | Yes |
| Lacustrine Deposit | Till | 27.48 | 0.0 | [23.33, 31.62] | Yes |

| Loess | Till | 4.23 | 0.0 | [2.48, 5.97] | Yes |

**Table 1.20 Multiple comparisons of yield mean from parent materials with Tukey's HSD, FWER= 0.01 in 2017**

| Group 1 | Group2 | Mean Diff. | p-adj | 99%CI | Reject |
|---|---|---|---|---|---|
| Lacustrine Deposit | Loess | 13.45 | 0.0 | [10.48, 16.42] | Yes |
| Lacustrine Deposit | Till | 16.41 | 0.0 | [13.56, 19.25] | Yes |
| Loess | Till | 2.96 | 0.0 | [1.75, 4.17] | Yes |

**Table 1.20 Multiple comparisons of yield mean from parent materials with Tukey's HSD, FWER= 0.01 in 2021**

| Group 1 | Group2 | Mean Diff. | p-adj | 99%CI | Reject |
|---|---|---|---|---|---|
| Lacustrine Deposit | Loess | 18.90 | 0.0 | [15.63, 22.18] | Yes |
| Lacustrine Deposit | Till | 22.37 | 0.0 | [19.22, 25.51] | Yes |
| Loess | Till | 3.47 | 0.0 | [2.14, 4.79] | Yes |

## 5.4.5 Yield analysis by Drainage classes

Soil drainage classes are inferences of the draining capability of soil based on soil color and composition. The yield in field 57 follows the expected decrease with a decrease in the draining capability of the soil, see Fig. 1.21. It is interesting to note here that the mean yield from SPD is lower than PD in 2021 albeit with minimal difference. A general monotonic trend should not be expected between yield and topographic variables. The VPD areas had significantly lower yields in all three available years. It had a specifically low yield in 2015 see 1.21(a) boxplot, whereas a comparatively low yield with the highest variation in (b) and (c) . However, the variation in yield in the VPD region is higher in 2017 and 2021. On more minute observation the areas of low yield in the VPD regions which have lower outcomes persistently in 2015, 2017, and 2021 also have lower elevations and higher slopes.  Indicating a low spot with VPD soil with lacustrine deposits as parent material and silty-clay-loam texture is a less yielding area in field 57. The yield-drainage class OLS regression analysis had low R-squared values which do not support a complete explanation of yield variation using the drainage class alone, however, the significant effect of the drainage classes on the yield is evident in the high F-statistic values and low probability of the F- statistics as seen in the results of ANOVA test results summarized in table 1.21, 1.22, and 1.23. The results of the post-hoc Tukey's HSD tests reveal the significance of the difference of average yields between SPD-VPD and PD-VPD.

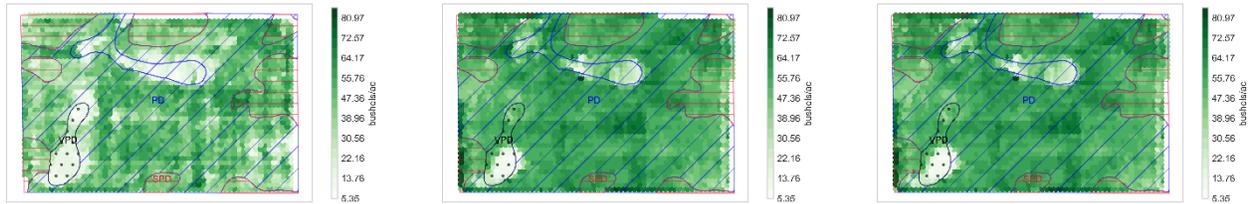

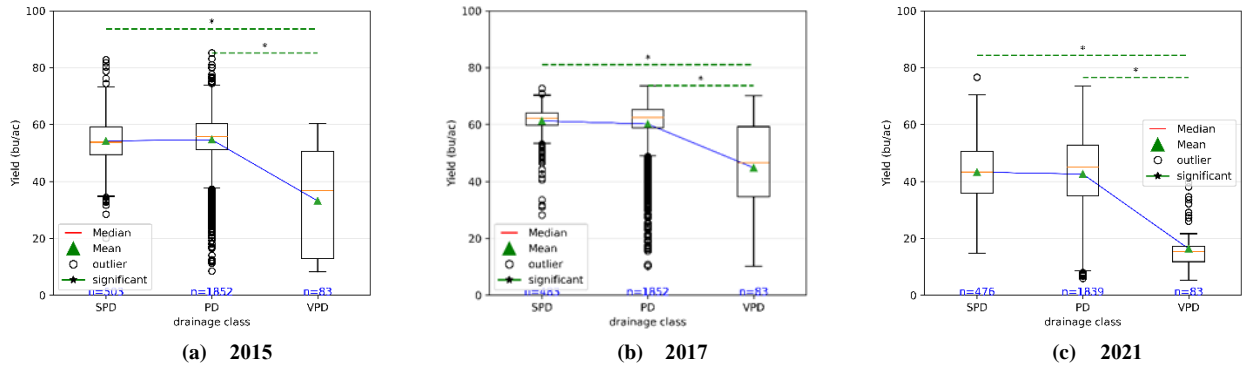

(a)  2015  (b)  2017  (c)  2021

**Fig. 1.21 Yield grouped by drainage class in field 57**

**Table 1.17 Analysis of variance of yield grouped by drainage class in 2015**

| Source | Sum of Squares | df | F-value | Pr(>F) |
|---|---|---|---|---|
| C(drainage class) | 2 ×10⁴ | 2 | 129.49 | <0.001 |
| Residual | 1.86 ×10⁵ | 2.42×10³ | - | - |

**Table 1.19 Analysis of variance of yield grouped by drainage class in 2021**

| Source | Sum of Squares | df | F-value | Pr(>F) |
|---|---|---|---|---|
| C(drainage class) | 3.68×10³ | 2 | 191.78 | <0.001 |
| Residual | 2.33×10⁵ | 2.44×10³ | - | - |

**Table 1.20 Multiple comparisons of yield mean from drainage classes with Tukey's HSD, FWER= 0.01 in 2015**

| Group 1 | Group2 | Mean Diff. | p-adj | 99%CI | Reject |
|---|---|---|---|---|---|
| PD | SPD | 0.69 | 0.54 | [-1.22, 2.61] | No |
| PD | VPD | -26.23 | 0.0 | [-30.41, -22.05] | Yes |
| SPD | VPD | -26.93 | 0.0 | [-31.36, -22.50] | Yes |

**Table 1.20 Multiple comparisons of yield mean from drainage classes with Tukey's HSD, FWER= 0.01 in 2017**

| Group 1 | Group2 | Mean Diff. | p-adj | 99%CI | Reject |
|---|---|---|---|---|---|
| PD | SPD | 1.11 | 0.04 | [-0.20, 2.42] | No |
| PD | VPD | -15.37 | 0.0 | [-18.24, -12.50] | Yes |
| SPD | VPD | -16.48 | 0.0 | [-19.52, -13.43] | Yes |

**Table 1.20 Multiple comparisons of yield mean from drainage classes with Tukey's HSD, FWER= 0.01 in 2021**

| Group 1 | Group2 | Mean Diff. | p-adj | 99%CI | Reject |
|---|---|---|---|---|---|
| PD | SPD | -0.41 | 0.68 | [-1.85, 1.02] | No |
| PD | VPD | -21.48 | 0.0 | [-24.69, -18.28] | Yes |
| SPD | VPD | -21.07 | 0.0 | [-24.45, -17.69] | Yes |

Notable observations while analyzing the effect of soil's physical properties on yield using descriptive analysis from box-and-whisker plots, analysis of variance in yield from four elevation groups, and post-hoc analysis results obtained using Tukey's HSD test on data from three years are noted below. The weather factors were included in the observations to create plausible inferences.

–The drainage class, texture, and parent material of soil significantly affect the average yield in their geospatial locations. Extreme weather patterns amplify these effects.

–The texture of soil has the lowest R-squared values among the selected soil properties followed by drainage class and parent material.

–area B was very poorly drained, with a silty clay loam texture, and had lacustrine deposits as the parent material. These properties make the soil hard to drain. Too much water holding capacity of the soil along with lower elevation and high slope makes B a low-yielding zone even with suitable weather patterns in field 57.

–Area marked by A is Poorly drained (PD), with silty clay loam texture, and has a loess parent material. This makes it better drained than B, however, it is at a lower elevation and higher slope. These factors cause lower yield in A, as seen in the yield maps. However, more information is required to ascertain the correct cause of this lower yield outcome in A.

–The ANOVA and Tukey's HSD test results indicate the OLS regression models were insufficient in explaining the yield variation based on a single topographic factor alone.

These observations in the analysis of infield yield variations support the use of geospatial blocking to minimize the variation by isolating the topographic bias from the natural variations.

# 6 Future Work

The proposed research will focus on developing an algorithm to generate geospatial blocks for classifying integrated landscape and soil properties. A key objective will be to ensure that these blocks effectively represent spatial heterogeneity, enabling robust classification using suitable machine-learning models. Various algorithms, such as Random Forest, Support Vector Machines, and Deep Learning methods, will be explored to determine the most efficient approach for classification. To validate the geospatial block approach, example field trials will be designed for Field 57, ensuring that experimental plots align with soil and landscape variations. The trials will provide data to refine the algorithm and improve its predictive accuracy. The WOFOST crop simulation model will be employed to evaluate these geospatially designed trials. Simulations will assess the impact of soil and landscape properties on crop performance, enabling predictive analysis of yield variability under different management strategies. Model calibration and validation will be carried out using observed data to enhance reliability. Future work will focus on optimizing geospatial block delineation, improving machine learning classification accuracy, and refining simulation parameters to enhance predictive capability. The integration of real-time sensor data and remote sensing inputs could further improve the dynamic adaptability of the model, making it a scalable tool for precision agriculture.

# REFERENCES

[1] W. B. Mercer and A. D. Hall, "The experimental error of field trials," The Journal of Agricultural Science, vol. 4, pp. 107–132, 1911.

[2] R. A. Fisher, Statistical methods for research workers. Oliver and Boyd, 1928.

[3] Fisher, Ronald Aylmer, The Design of Experiments. Hafner publishing company INC., 1935.

[4] F. Yates, "Sir ronald fisher and the design of experiments," pp. 307–321, 1964. [Online]. Available: https://about.jstor.org/terms

[5] A. Borges, A. González-Reymundez, O. Ernst, M. Cadenazzi, J. Terra, and L. Gutiérrez, "Can spatial modeling substitute for experimental design in agri- cultural experiments?" Crop Science, vol. 59, pp. 44–53, 1 2019.

[6] R. Hoefler, P. González-Barrios, M. Bhatta, J. A. Nunes, I. Berro, R. S. Nalin, A. Borges, E. Covarrubias, L. Diaz-Garcia, M. Quincke, and L. Gutierrez, "Do spatial designs outperform classic experimental designs?" Journal of Agricul- tural, Biological, and Environmental Statistics, vol. 25, pp. 523–552, 12 2020.

[7] M. D. Casler, "Fundamentals of experimental design: Guidelines for designing successful experiments," Agronomy Journal, vol. 107, pp. 692–705, 3 2015.

[8] H. M. van Es, C. P. Gomes, M. Sellmann, and C. L. van Es, "Spatially-balanced complete block designs for field experiments," Geoderma, vol. 140, pp. 346–352, 8 2007.

[9] R. Khosla, D. G. Westfall, R. M. Reich, J. S. Mahal, and W. J. Gangloff, "Geo- statistical applications for precision agriculture," Geostatistical Applications for Precision Agriculture, 2010.

[10] J. A. King, P. M. R. Dampney, R. M. Lark, H. C. Wheeler, R. I. Bradley, and T. R. Mayr, "Mapping potential crop management zones within fields: Use of yield-map series and patterns of soil physical properties identified by electromag- netic induction sensing," 2005.

[11] J. Galamboŝová, V. Rataj, Prokeinová, and J. Preŝinská, "Determining the man- agement zones with hierarchic and non-hierarchic clustering methods," Res. Agr. Eng., vol. 60, 2014.

[12] A. Layton, J. V. Krogmeier, A. Ault, and D. R. Buckmaster, "From yield history to productivity zone identification with hidden markov random fields," Precision Agriculture, vol. 21, pp. 762–781, 2020. [Online]. Available: https://doi.org/10.1007/s11119-019-09694-2

[13] M. S. Cox and P. D. Gerard, "Soil management zone determination by yield stability analysis and classification," Agronomy Journal, vol. 99, pp. 1357–1365, 9 2007.

[14] R. A. Ortega and O. A. Santibáñez, "Determination of management zones in corn (zea mays l.) based on soil fertility," Computers and Electronics in Agriculture, vol. 58, pp. 49–59, 8 2007.

[15] K. L. Fleming, D. G. Westfall, D. W. Wiens, and M. C. Brodahl, "Evaluating farmer defined management zone maps for variable rate fertilizer application," 2012.

[16] J. A. Taylor, B. Tisseyre, and C. Leroux, "A simple index to determine if within- field spatial production variation exhibits potential management effects: appli- cation in vineyards using yield monitor data," Precision Agriculture, vol. 20, pp. 880–895, 10 2019.

[17] E. Scudiero, P. Teatini, D. L. Corwin, R. Deiana, A. Berti, and F. Morari, "Delineation of site-specific management units in a saline region at the venice lagoon margin, italy, using soil reflectance and apparent electrical conductivity," Computers and Electronics in Agriculture, vol. 99, pp. 54–64, 2013.

[18] Q. Zhu, J. P. Schmidt, and R. B. Bryant, "Maize (zea mays l.) yield response to nitrogen as influenced by spatio-temporal variations of soil-water-topography dynamics," Soil and Tillage Research, vol. 146, pp. 174–183, 3 2015.

[19] X. Huang, L. Wang, L. Yang, and A. N. Kravchenko, "Management effects on relationships of crop yields with

topography represented by wetness index and precipitation," Agronomy Journal, vol. 100, pp. 1463–1471, 9 2008.

[20] B. R. Khakural, P. C. Robert, and D. J. Mulla, Relating Corn/Soybean Yield to Variability in Soil and Landscape Characteristics, 11 2015, pp. 117–128.

[21] S. J. Leuthold, O. Wendroth, M. Salmer´on, and H. Poffenbarger, "Weather- dependent relationships between topographic variables and yield of maize and soybean," Field Crops Research, vol. 276, 2 2022.

[22] S. J. Leuthold, M. Salmer´on, O. Wendroth, and H. Poffenbarger, "Cover crops decrease maize yield variability in sloping landscapes through increased water during reproductive stages," Field Crops Research, vol. 265, 5 2021.

[23] S. Peukert, R. Bol, W. Roberts, C. J. Macleod, P. J. Murray, E. R. Dixon, and
R. E. Brazier, "Understanding spatial variability of soil properties: A key step in establishing field- to farm-scale agro-ecosystem experiments," Rapid Commu- nications in Mass Spectrometry, vol. 26, pp. 2413–2421, 10 2012.

[24] P. Jiang and K. D. Thelen, "Effect of soil and topographic properties on crop yield in a north-central corn-soybean cropping system," Agronomy Journal, vol. 96,
pp. 252–258, 2004.

[25] L. Guo, Y. Yang, Y. Zhao, Y. Li, Y. Sui, C. Tang, J. Jin, and X. Liu, "Reducing topsoil depth decreases the yield and nutrient uptake of maize and soybean grown in a glacial till," Land Degradation and Development, vol. 32, pp. 2849–2860, 5 2021.

[26] A. N. Kravchenko, G. P. Robertson, K. D. Thelen, and R. R. Harwood, "Man- agement, topographical, and weather effects on spatial variability of crop grain yields," Agronomy Journal, vol. 97, pp. 514–523, 2005.

[27] A. N. Kravchenko and D. G. Bullock, "Correlation of corn and soybean grain yield with topography and soil properties," Agronomy Journal, vol. 92, pp. 75–83, 2000.

[28] S. R. Rahmani, Z. Libohova, J. P. Ackerson, and D. G. Schulze, "Estimating natural soil drainage classes in the wisconsin till plain of the midwestern u.s.a. based on lidar derived terrain indices: Evaluating prediction accuracy of multi- nomial logistic regression and machine learning algorithms," Geoderma Regional, vol. 35, 12 2023.

[29] Soil Survey Staff, Natural Resources Conservation Service, United States De- partment of Agriculture, "Soil survey geographic (ssurgo) database," https:
//sdmdataaccess.sc.egov.usda.gov, accessed: [5/30/2024].

[30] Soil Survey Staff, "Soil survey of tippecanoe county, indiana," 1998. [Online].
Available: https://archive.org/details/tippecanoeIN1998

[31] "Official series description - chalmers series." 2011, accessed: [5/30/2024]. [Online]. Available: https://soilseries.sc.egov.usda.gov/OSD Docs/ C/CHALMERS.html

[32] Soil Survey Staff, Pella, "Official series description - pella series." 2011, accessed: [5/30/2024]. [Online]. Available: https://soilseries.sc.egov.usda.gov/OSD Docs/ P/PELLA.html

[33] Soil Survey Staff,Milford, "Official series description - milford series." 2011, accessed: [5/30/2024]. [Online]. Available: https://soilseries.sc.egov.usda.gov/ OSD Docs/P/MILFORD.html

[34] Soil Survey Staff, Brenton, "Official series description - brenton series." 2011, accessed: [5/30/2024]. [Online]. Available: https://soilseries.sc.egov.usda.gov/ OSD Docs/P/BRENTON.html

[35] U2U, "Useful to usable (u2u) com groups purduegdd,"2017,accessed: [5/30/2024]. [Online]. Available: https://mygeohub.org/groups/u2u/purdue gdd#map=9/-86.89856/40.43247

[36] R. Lark, J. Stafford, and H. Bolam, "Limitations on the spatial resolution of yield mapping for combinable crops," Journal of Agricultural Engineering Research, vol. 66, no. 3, pp. 183–193, 1997.